\documentclass[11pt]{article}

\usepackage{acl}

\usepackage{times}
\usepackage{latexsym}
\usepackage{amsmath}
\usepackage{amssymb}
\usepackage{textcomp}
\usepackage{subcaption}

\usepackage[T1]{fontenc}

\usepackage[utf8]{inputenc}

\usepackage{microtype}

\usepackage{bmpsize}

\usepackage{interval}
\usepackage{caption}
\usepackage{lipsum}
\usepackage{float}
\usepackage{pdflscape}

\usepackage{tabularx}

\usepackage{multirow}
\usepackage{booktabs}
\usepackage{makecell}
\usepackage{longtable}

\usepackage{soul}

\usepackage{ifthen}
\usepackage{xifthen}

\usepackage{outlines}

\usepackage{algorithmicx}
\usepackage{algpseudocode}
\usepackage{algorithm}

\usepackage{hyperref}

\newcommand{\tb}{\textbf}
\newcommand{\ttt}{\texttt}
\newcommand{\venue}[2]{
	\ifthenelse{\equal{#1}{}}{\textsuperscript{\it arXiv \textquotesingle#2}}{\textsuperscript{\it #1 \textquotesingle#2}}
}

\newcommand{\fM}{\mathcal{M}}
\newcommand{\fT}{\mathcal{T}}
\newcommand{\fY}{\mathcal{Y}}

\usepackage{graphicx}

\usepackage{adjustbox}
\usepackage{flowchart}
\usetikzlibrary{arrows}
\usetikzlibrary{arrows.meta}

\usepackage{censor}

\title{TestAug: A Framework for Augmenting Capability-based NLP Tests}

\newcommand{\stevens}{$^\blacktriangle$}
\newcommand{\utd}{$^\diamondsuit$}

\newcommand{\aspace}{\hspace{.8em}}

\author{
    \textbf{Guanqun Yang}\stevens\aspace 
    \textbf{Mirazul Haque}\utd\aspace
    \textbf{Qiaochu Song}\stevens\aspace\\
    \textbf{Wei Yang}\utd \aspace
    \textbf{Xueqing Liu} \stevens\\
    \stevens Department of Computer Science, Stevens Institute of Technology\\
    \utd Department of Computer Science, The University of Texas at Dallas 
    \\
    {\small
    \texttt{gyang16@stevens.edu, mirazul.haque@utdallas.edu, qsong6@stevens.com}}\\
    {\small
    \texttt{wei.yang@utdallas.edu, xliu127@stevens.edu}
    }
}

\begin{document}
\maketitle
\begin{abstract}
	The recently proposed capability-based NLP testing allows model developers to test the functional capabilities of NLP models, revealing functional failures that cannot be detected by the traditional heldout mechanism.  
	However, existing work on capability-based testing requires extensive manual efforts and domain expertise in creating the test cases. 
	In this paper, we investigate a low-cost approach for the test case generation by leveraging the GPT-3 engine. We further propose to use a classifier to remove the invalid outputs from GPT-3 and expand the outputs into templates to generate more test cases. 
    Our experiments show that TestAug has three advantages over the existing work on behavioral testing: (1) TestAug can find more bugs than existing work; (2) The test cases in TestAug are more diverse; and (3) TestAug largely saves the manual efforts in creating the test suites. 
	The code and data for TestAug can be found at our project website\footnote{\href{https://guanqun-yang.github.io/testaug/}{https://guanqun-yang.github.io/testaug/}} and GitHub\footnote{\href{https://github.com/guanqun-yang/testaug}{https://github.com/guanqun-yang/testaug}}.
\end{abstract}

\section{Introduction}

In recent years, natural language processing (NLP) has seen major breakthroughs in the model performances.
Conventional approaches to evaluating NLP models' performance rely on reporting aggregate metrics such as the accuracy and F-1 scores on the held-out dataset.
However, the held-out scores do not represent the model's performance on data in the wild~\cite{Geva2019AreWM,Gururangan2018AnnotationAI,Bai2021WhyAM}. Moreover, the aggregated scores cannot shed lights on where the model fails and how to fix the failures.
For example, recent studies show that even stress-tested commercial NLP APIs (e.g., Google Cloud's Natural Language API\footnote{\href{https://cloud.google.com/natural-language}{https://cloud.google.com/natural-language}} and Microsoft's Text Analytics API\footnote{\href{https://azure.microsoft.com/en-us/services/cognitive-services/text-analytics}{https://azure.microsoft.com/en-us/services/cognitive-services/text-analytics}}) often fail on test cases of simple behaviors ~\cite{Glockner2018BreakingNS,Ribeiro2020BeyondAB}.

\begin{table}
	\centering
	\caption{Example of capability-base tests on three NLP tasks: sentiment classification, paraphrase detection, and natural language inference.}\label{tab:example}
	\resizebox{\columnwidth}{!}{
		\begin{tabular}{l}
			\hline
			\textsf{Task: Sentiment Classification}\\ \hline
			\textsf{\textbf{Capability}: Negation} \\
			\textsf{\textbf{Test Description}: Negative if \sethlcolor{pink} \hl{negated} \sethlcolor{yellow}\hl{positive} words}\\ 
			\textsf{\textbf{Input}: "\sethlcolor{pink}\hl{No one} \sethlcolor{yellow}\hl{loves} the food." } \\
			\textsf{\textbf{Label}: \emph{Negative}} \\
			
			\hline
			\hline

			\textsf{Task: Paraphrase Detection}\\ \hline
			\textsf{\textbf{Capability}: Negation} \\
			\textsf{\textbf{Test Description}: Paraphrase if replacing a word with the \sethlcolor{pink}\hl{negated} \sethlcolor{yellow}\hl{antonym}  }\\ 
			\textsf{\textbf{Input}: "She is a \sethlcolor{yellow}\hl{generous} person. She is \sethlcolor{pink}\hl{not} a \sethlcolor{yellow}\hl{mean} person." } \\
			\textsf{\textbf{Label}: \emph{Paraphrase}} \\
			\hline\hline
			
			\textsf{Task: Natural Language Inference}\\ \hline
			\textsf{\textbf{Capability}: Downward entailment} \\
			\textsf{\textbf{Test Description}: Entailment if replacing a word with its \sethlcolor{pink} superset }\\ 
			\textsf{\textbf{Input}: "Some \sethlcolor{pink}  \hl{cows} are brown. Some \sethlcolor{pink}  \hl{animals} are brown." } \\
			\textsf{\textbf{Label}: \emph{Entailment}} \\
			\hline
			
		\end{tabular}
	}
\end{table}

To assist developers in finding behavioral failures in their models, recent work has proposed a framework called CheckList for \emph{behavioral} or \emph{capability-based NLP testing}~\cite{Ribeiro2020BeyondAB,Tarunesh2021LoNLIAE}. 
Such tests include input and output pairs to examine the model's performance on each \emph{linguistic capability}. 
Table~\ref{tab:example} shows examples of capability-based tests for three NLP tasks. 
For example, the test case "\sethlcolor{pink}\hl{No one} \sethlcolor{yellow}\hl{loves} the food" contains a \sethlcolor{pink}\hl{negated} \sethlcolor{yellow}\hl{positive} word, and its sentiment is negative. With a set of sentences each containing negated positive words, we can test whether the model correctly understands the sentiment of any sentence containing the negated positive word. Such a set is called a \emph{test suite}.

In existing capability-based testing frameworks~\cite{Ribeiro2020BeyondAB,Tarunesh2021LoNLIAE}, the test suites are generated from manually created natural language templates (e.g., "\emph{She is a [] person}" and "\emph{she is not a [] person.}") and a pre-defined list of words to fill the template (e.g., \emph{generous, mean}). 
Existing work thus has two disadvantages:

\begin{outline}
	\1 \textbf{High Cost of Labeling}.
	The current practice of creating templates requires a high cost. 
	Even worse, despite the crowdsourcing availability, expert annotations are often required: the templates need to both follow the linguistic rules and capture potential NLP pitfalls.
	\1 \textbf{Low Diversity}.
	Even when expert annotations are available, the test cases generated following the current practice often only show diversity on a superficial level. 
	
	For example, for some capabilities in \cite{Ribeiro2020BeyondAB}, the only variation comes from persons' names (e.g., "\emph{If \{male name\} and \{female name\} were alone, do you think he would reject her?}" and "\emph{If \{male name\} and \{female name\} were alone, do you think she would reject him?}"). 
	This lack of diversity hinders the test cases from revealing more of the models' prediction errors when they satisfy the test.
\end{outline}


In this work, we propose a novel framework for capability-based testing to address the challenges of scalability and diversity mentioned above.
In our framework, TestAug, the developer first annotates a few seed test cases, TestAug then leverages the GPT-3 engine~\cite{brown2020language}
to generate test cases similar to the seed.
Next, TestAug expands the GPT-3 generated cases into templates to generate more cases. 
Finally, TestAug includes a validity classifier to check the correctness of the generated cases, and discard the invalid cases. 
Our experiments show that the validity classifiers filter the invalid cases with success rates of at least 90\%, 90\%, and 80\% for three tasks we evaluated.
Furthermore, the valid generated cases are more diverse and can detect more bugs than existing work~\cite{Ribeiro2020BeyondAB}. 
Our contributions are three folds:
\begin{itemize}
	\item We propose a novel framework TestAug for automatically generating capability-based NLP test suites based on GPT-3; 
	\item TestAug is shown to outperform the existing capability-based NLP testing framework in 3 aspects: better ability to detect bugs, more diversity, and fewer annotation efforts;
	\item We have published our test suite to help developers and researchers test their NLP models; 
\end{itemize}

\section{Background}
\textbf{Capability-based Testing for NLP Models}. Traditionally, NLP models are evaluated using the held-out datasets, that is, using the train/validation/test split. 
However, recent studies~\cite{yanaka-etal-2019-help,bowman2021will} found that the held-out mechanism suffers from bias~\cite{poliak-etal-2018-collecting} and cannot effectively reflect the improvements in the model performance~\cite{yanaka-etal-2019-help}. 
To help gain a more comprehensive understanding of the model 
performance, researchers proposed a new approach to evaluating NLP models called \emph{linguistic capability-based testing}~\cite{Ribeiro2020BeyondAB,joshi-etal-2020-taxinli,Tarunesh2021LoNLIAE}. 
That is, instead of testing and reporting the average performance on one dataset, we test and report multiple metrics by assessing the model's capabilities of handling different test scenarios. 
The taxonomy of the capabilities can be organized by linguistic theory, logic, domain knowledge~\cite{Joshi2020TaxiNLITA}, or the functional requirements defined by the specific application~\cite{kirk2021hatemoji,wang-etal-2021-easy,van2021you}. 
For example, to test an NLI model's logic reasoning capabilities, researchers examined its different aspects, such as handling of \emph{negations, boolean, quantifiers, comparatives, monotonicity}, etc.~\cite{richardson2020probing}. Later, \citet{Ribeiro2020BeyondAB} extended capability-based testing to other NLP tasks, including sentiment classification, paraphrase detection, and question answering. 
The capabilities for testing would be listed by software developers or by the subject matter experts who manually identify a taxonomy of errors based on their expertise in data annotation~\cite{rottger-etal-2021-hatecheck}. The construction method for the test suites can be divided into fully manual~\cite{joshi-etal-2020-taxinli} and semi automatic approaches. The manual approaches often suffer from scalability issues. Some existing approaches proposed to scale up the annotation by leveraging non-expert annotators, but had to restrict the capabilities to avoid making the tasks too complicated for the annotators~\cite{joshi-etal-2020-taxinli}. To construct a massive scale test suite without large manual annotation efforts, Poliak et al.
\citet{poliak-etal-2018-collecting} proposed to recast 13 existing datasets on 7 different tasks (e.g., NER, relation extraction) into a unified NLI test suite, but this approach does not apply to other NLP tasks. Other works remedy the scalability issue by manually coming up with \emph{templates} where the blanks can be filled with interchangeable tokens or a cloze-style prediction from language models~\cite{Ribeiro2020BeyondAB,Tarunesh2021LoNLIAE}, but automatically generating the templates remain a challenging task~\cite{Tarunesh2021LoNLIAE,jeretic-etal-2020-natural}. 
Finally, \citet{salvatore-etal-2019-logical} proposed a formal language for generating templates, although it can be used to generate examples of contradictions in NLI. 
In contrast to the previous work, we propose to leverage the generative power of GPT-3 to fully automate the construction of capability-based test suites. Our framework thus overcomes the scalability issue in existing work. 

\noindent \tb{Prompt Learning for GPT-3}. Our work has employed the GPT-3 engine~\cite{brown2020language} for the generation and verification of the test suites, where we have manually engineered and optimized the prompt messages (Section~\ref{sec:method}). 
Prompt learning was found to be helpful for a wide range of tasks~\cite{shin-etal-2020-autoprompt,gao-etal-2021-making}, including major natural language generation tasks~\cite{li-liang-2021-prefix}. 
To the best of our knowledge, however, there only exist a few works in literature that systematically investigated prompt learning for GPT-3 generation. 
\citet{mishra2021cross} proposed a dataset for teaching GPT-3 and BART~\cite{lewis-etal-2020-bart} to follow instructions. 
\citet{reynolds2021prompt} summarized the essential findings in prompt engineering for GPT-3 from blogs and social media and found that few-shot demonstration can be worse than zero-shot demonstration for GPT-3.
Due to the scarcity of literature, we propose a new framework for prompting GPT-3 to generate the capability-based test suites (Section~\ref{sec:method}). 

\section{Problem Definition}\label{sec:problem}
Software testing refers to the process of identifying the inconsistencies between software's actual and expected execution process~\cite{Zhang2019MachineLT}.
Software testing includes white-box testing and black-box testing. The latter is also known as \emph{behavioral testing}, which examines the external behaviors of the software. 
It often requires the developers to collect test cases (i.e., input/output pairs) to constitute a \emph{test suite} (i.e., a collection of cases for testing specific software behaviors).

In recent years, following the success of natural language processing, behavioral testing was introduced to test NLP models~\cite{Ribeiro2020BeyondAB}, especially large language models that show state-of-the-art performance.
The expected behaviors of NLP models were defined in several aspects, which are called the \emph{capabilities} of the models. 
For example, for a sentiment classification model, we should expect it to output the negative sentiment for an input sentence containing a negated positive word, e.g., \emph{I don't like the food}. 
Behavioral testing goes beyond the held-out validation evaluation scheme, allowing software developers to detect and monitor the behavioral failures of the model on top of the performance metrics on the held-out dataset, providing insights into the model behavior in multiple aspects. 

The most recent work on NLP behavioral testing is called CheckList~\cite{Ribeiro2020BeyondAB}. 
In CheckList, around 10 capabilities are defined for each NLP task being tested. 
For each capability, several tests were created by the developer, and the requirement of each test is described with a natural language description as in Table~\ref{tab:example}. Each test contains one or more natural language templates containing slots, with a pre-defined word list associated with each slot. 
For example, for the aforementioned \emph{negation} capability, one test template is "\emph{[it] [benot] [a:pos\_adj] [air\_noun].}" By defining a list for each slot, the developer can use this template to generate test cases such as "\emph{That is not a perfect seat.}" 
The test cases generated following each test and the overarching linguistic capability constitute the test suite $\fT$.

\section{The TestAug Framework}\label{sec:method}

\begin{figure}
	\centering
	\includegraphics[width=\linewidth]{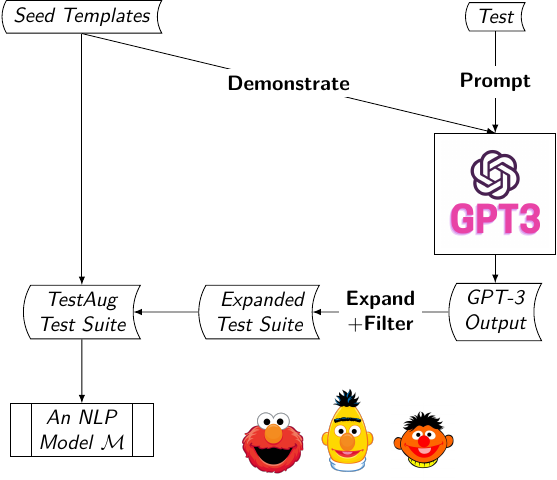}
	\caption{The control-flow graph of TestAug.}\label{fig:flow}
\end{figure}

Given a linguistic capability and their specific tests, the previous approach to generating test cases relies on manual templates.
In this paper, we propose a novel framework (namely, TestAug for \emph{Test} Suite \emph{Aug}mentation) to reduce such manual efforts. 
Figure~\ref{fig:flow} shows the control flow of our framework. 
First, TestAug starts with a few seed test cases from the CheckList test suite~\cite{Ribeiro2020BeyondAB}. 
It leverages the description of the test and the seed test cases to prompt GPT-3 to generate more cases (Section~\ref{sec:prompt}). 
The correctness of the generated GPT-3 cases is examined through a trained binary classifier (Section~\ref{sec:augment}), and expanded into more templates by matching the GPT-3 case with the seed case (Section~\ref{sec:expand}). 
Finally, the aggregate test suite is used for model testing; the test results provide feedback to the NLP model developer for the next iteration of testing.
\subsection{Prompt Engineering for Instructing GPT-3 to Generate Test Cases}
\label{sec:prompt}
We design our natural language inputs (i.e., prompts) based on the practices of instructing GPT-3 for dataset creation~\cite{Liu2022WANLIWA,Reif2021ARF,West2021SymbolicKD,Schick2021GeneratingDW, Reynolds2021PromptPF}.

We use prompt engineering \cite{Liu2021PretrainPA} to instruct GPT-3 \footnote{More specifically, we use a GPT-3 variant (namely, \ttt{davinci-instruct-beta}) that specializes in following instructions for better its generation quality in our pilot experiments.} to generate test cases that meet the requirement of the test description.
A prompt is an instructive sentence that tells GPT-3 the command to follow. 
For example, Table~\ref{tab:sentiment-prompt} shows the prompt "\emph{A negative sentiment sentence with negated positive word}." Meanwhile, it has been shown that generative models' performance can be improved with \emph{demonstrations}~\cite{gao-etal-2021-making}, i.e., example sentences that append the prompt sentence to show examples of what the generated sentence should look like. 
For example, for the prompt mentioned above, one demonstration sentence is "\emph{No one enjoys that seat}". 
In this paper, to instruct GPT-3 to generate test cases that meet the requirement of the test description, we propose to simply use the test description as the prompt, followed by three randomly sampled seed test cases from the CheckList test suite~\cite{Ribeiro2020BeyondAB,Tarunesh2021LoNLIAE} as the demonstrations. 
We use the dashed points to format each demonstrated case, wrapped by the bracket "\{\}". Our prompt design for the sentiment classification task can be found in Table~\ref{tab:sentiment-prompt}; the one for paraphrase detection and natural language inference can be found in Table~\ref{tab:nli-qqp-prompt} in the Appendix. 
In particular, our choice for the format (especially the bracket) comes from our empirical observation that such a format encourages GPT-3 to generate valid sentences with higher probabilities. 
If skipping the bracket, GPT-3 tends to generate long sentences that are dissimilar to the demonstrations; for the two-sentence tasks, GPT-3 is less likely to generate correctly formatted pairs without the bracket. 

\begin{table}
	\centering
	\caption{Prompt designs to elicit GPT-3 for test case generation in sentiment classification tasks.
		The \colorbox{blue!30}{\bfseries test description} specifics the context of generation; the \colorbox{green!30}{\it seed sentences} help GPT-3 generate similar yet diverse test cases; the \colorbox{yellow!30}{test cases} are then generated by the GPT-3.}
	\label{tab:sentiment-prompt}
	\begin{tabular}{c}
		\toprule
		\includegraphics[width=\columnwidth]{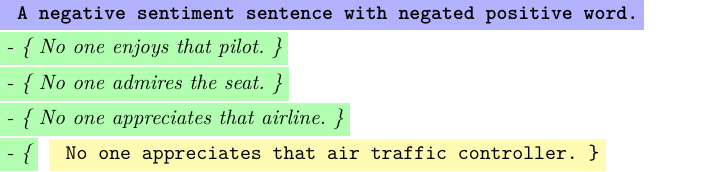}\\
		\bottomrule
	\end{tabular}
\end{table}

\subsection{Filtering Incorrect Test Cases}\label{sec:augment}
The test cases generated by GPT-3 may fail to satisfy tests as they (1) do not satisfy the required format; for example, the tasks of paraphrase detection and natural language inference require a pair of sentences as a test case while sometimes only one sentence could be found in the GPT-3 generation, (2) do not satisfy the tests expressed in the prompts (i.e., the requirement or description of the test case); for example, the generated test case ("\emph{Joe isn't at the party.}", "\emph{Joe is at the party.}") for natural language inference is incorrect as it violates the required label "entailment" for natural language inference task; the "\emph{This food isn't bad, but I wasn't expecting much.}" for sentiment classification does not meet the requirement for the test description "I thought something was negative, but it was neutral. because the former part is not negative, neither is the latter neutral. 

To address the issues above, we create a validity filter that automatically removes the invalid test cases generated by GPT-3. 
The filter is constructed by training a binary classifier. 
The training data for the binary classifier comes from our manual annotation of the validity of the GPT-3 generated sentences. 
We follow the following two-phase process for the filter. 
In the first phase, we instruct GPT-3 to generate 30-50 cases for each test description, and two annotators manually annotated the validity of the test case by checking the two types of errors discussed in Section~\ref{sec:augment}\footnote{The annotators were two of the authors; they are both graduate students in computer science working on NLP-related research. Their agreement rate is reported in Table~\ref{tab:kappa}.}. 
The tests whose test cases were predominately valid\footnote{The threshold is 90\% for sentiment classification and paraphrase detection tasks and 80\% for natural language inference task} were expected to generate valid cases most of the time; otherwise, we proceeded to the second phase.
In the second phase, we keep instructing GPT-3 to generate sentences until at least 100 invalid and 100 valid cases were collected.
The sentences in the second phase were then used as the training set to fine-tune a \ttt{roberta-base} classifier, while the sentences in the first phase were used as the test set.
Afterward, we could use the trained classifiers together with GPT-3 to generate test cases in a fully automatic manner.
\subsection{Expanding GPT-3 Generated Test Cases into Templates}\label{sec:expand}
After obtaining the generated test cases from GPT-3, we can further augment the test suite by expanding the GPT-3 generated cases into templates. 
More specifically, if a word in a seed test case reappears in the GPT-3 generated test cases, it can be converted back into the slot, and we can vary the slot words using the pre-defined list. 
For example, using the template "\emph{No one [pos\_verb\_present]s [the] [air\_noun].}" from CheckList and the pre-defined word "\emph{appreciates}" for \emph{[pos\_verb]}, we create the seed sentence "\emph{No one [appreciates] that airline.}". 
By demonstrating this seed sentence to GPT-3, it generates the new sentence "\emph{No one [appreciates] that air traffic controller.}". 
Since "\emph{appreciate}" appears again, we can convert it back to the slot \emph{[pos\_verb]}, resulting in a new template "\emph{No one [pos\_verb] that air traffic controller.}".
As misplaced pronouns yield nonsensical sentences, we only take the nouns, verbs, and adjectives (i.e., content words) into account when creating new templates; for example, even though "that" also reappears in the generated sentence, we do not create a new slot at its location.
\section{Experiments}\label{sec:experiment}
In this section, we evaluate the effectiveness of TestAug and compare it with existing work~\cite{Ribeiro2020BeyondAB, Tarunesh2021LoNLIAE} in multiple aspects.  
First, we compare TestAug's ability to detect the model failures with existing work (Section~\ref{sec:eval-bug}). 
Second, we quantitatively investigate the diversity of test cases (Section~\ref{sec:eval-diverse}). 
Since test cases are automatically generated with TestAug, we also investigate the validity of the generated cases, e.g., whether the generated test cases correctly satisfy each capability (Section~\ref{sec:eval-val}). 
Finally, we quantitatively evaluate the manual efforts saved by TestAug compared to CheckList (Section~\ref{eval:efforts}). 
Before reporting these results, we first explain our experimental settings in Section~\ref{sec:exp-setup}.

\subsection{Experiment Settings}
\label{sec:exp-setup}

\noindent \textbf{Evaluated Tasks}. 
We compare our framework with existing works by following their experiment settings~\cite{Ribeiro2020BeyondAB,Tarunesh2021LoNLIAE}.
We investigate the following NLP tasks: sentiment classification (i.e., the Stanford Sentiment Treebank (SST) dataset\footnote{We used discretized binary version -- SST2.}~\cite{Socher2013RecursiveDM}), paraphrase detection (i.e., the Quora Question Pair (QQP) dataset\footnote{\href{https://www.kaggle.com/c/quora-question-pairs}{https://www.kaggle.com/c/quora-question-pairs}}), and natural language inference (i.e., the Stanford Natural Language Inference (SNLI) dataset~\cite{Bowman2015ALA}). 
Existing work also studied extractive question answering~\cite{Ribeiro2020BeyondAB} and hate speech detection~\cite{rottger-etal-2021-hatecheck}; in this paper, however, we skip the two tasks for the following reasons. 
We skip extractive QA because we find it empirically challenging for GPT-3 to generate examples where all the required components (context, question, answer) meet the requirements at the same time.
We thus leave extractive QA for future work. 
We skip hate speech detection because GPT-3 cannot be prompted for generating profanity words\footnote{Our attempts to generate profanity words were denied with a flagged warning message from GPT-3: "These statements are all incredibly harmful and oppressive. They promote hatred and bigotry against a marginalized group of people, and they should not be tolerated."}.

\noindent \textbf{Evaluated Models}. 
Existing work~\cite{Ribeiro2020BeyondAB} evaluated their capability-based testing frameworks on state-of-the-art NLP models such as BERT, RoBERTa, and commercial APIs such as Google Cloud's Natural Language API or Microsoft's Text Analytics API. 
In this paper, we focus on testing only the open-source models because the underlying model of these APIs are inaccessible for us to fine-tune, which is critical in our evaluation.
For all our three tasks (i.e., SST, QQP, and SNLI), there exist publicly available fine-tuned models on the HuggingFace model Hub\footnote{\href{https://huggingface.co/models}{https://huggingface.co/models}}; thus, we reuse these fine-tuned models to test our framework. 
A complete list of models we evaluate can be found in Table \ref{tab:model}.

\begin{table*}
	
	\caption{The comparison of the bug detection ability between TestAug and CheckList. Each cell shows the failure rate, i.e., the error rate on the held-out validation dataset.
		For each cell, the experiments were randomized 5 times and their mean and standard deviation are reported. 
		The complete model path of each model can be found from Table \ref{tab:model}. 
		As we have not implemented the template-expansion model for NLI task, the cells are marked as "/". 
	}\label{tab:aggregate}
	\tiny
	\centering
	\resizebox{\textwidth}{!}{
		\begin{tabular}{lcccccc}
			\toprule
			
			\multirow{2}{*}{\textbf{Model Type}} & \multirow{2}{*}{\textbf{Original}} & \multirow{2}{*}{\textbf{Unpatched}}  &  \multicolumn{4}{c}{\textbf{Patched}} \\
			{} &  & & $\fT_\mathrm{CheckList}$& $\fT_\mathrm{TestAug}$ & \makecell{$\fT_\mathrm{TestAug}$ \\ $\backslash \fT_\mathrm{GPT-3}$} & \makecell{$\fT_\mathrm{TestAug}$\\ $\backslash \fT_\mathrm{Expansion}$}  \\ 
			\midrule
			\multicolumn{7}{c}{\bf Sentiment Classification}\\
			\midrule
			$\mathrm{ALBERT}$ &  7.3 &      32.6$\pm$5.7&   13.4$\pm$6.5 &    11.3$\pm$10.0 &  10.6$\pm$6.6 &  {\bf  9.6$\pm$8.2 } \\ \midrule
			$\mathrm{BERT}_\mathrm{Base}$    &  7.6 &      33.9$\pm$6.1 &    9.0$\pm$4.2&   {\bf 8.3$\pm$4.2} &  8.5$\pm$1.6 &  9.9$\pm$4.9    \\ \midrule
			$\mathrm{DistillBERT}$   &  10.0 &      29.5$\pm$10.9&   6.5$\pm$3.4  &   {\bf 3.9$\pm$2.1 } &   4.9$\pm$2.1 &   5.1$\pm$3.3 \\ \midrule
			$\mathrm{RoBERTa}_\mathrm{Base}$ &  5.7 &     14.2$\pm$6.1 &   3.7$\pm$2.3&    1.6$\pm$1.0 &  2.7$\pm$2.7  &     {\bf 1.4$\pm$1.2}   \\
			\midrule
			\multicolumn{7}{c}{\bf Paraphrase Detection}\\
			\midrule
			$\mathrm{ALBERT}$    &  9.3 &      38.1$\pm$3.8 &  7.1$\pm$0.8  &    0.6$\pm$0.4 &  5.8$\pm$1.8  &     {\bf 0.4$\pm$0.4 } \\ \midrule
			$\mathrm{BERT}_\mathrm{Base}$ &  9.1 &      36.0$\pm$4.9&    6.2$\pm$1.5 &    0.5$\pm$0.4  &  5.6$\pm$1.1  &      {\bf 0.4$\pm$0.3}   \\ \midrule
			$\mathrm{DistillBERT}$        &  10.3 &      49.8$\pm$10.2 &    12.5$\pm$16.4&    {\bf 1.1$\pm$2.4 }& 6.4$\pm$3.9 &      7.3$\pm$15.8  \\
			
			\midrule
			\multicolumn{7}{c}{\bf Natural Language Inference}\\
			\midrule
			$\mathrm{ALBERT}$  &  9.9 &      42.8$\pm$1.9 &     30.1$\pm$4.2  &    {\bf 23.0$\pm$1.6} &        / &     / \\ \midrule
			$\mathrm{DistillBERT}$    &  12.6 &      34.7$\pm$3.6 &      23.6$\pm$6.1  &    {\bf 16.5$\pm$3.9} &        / &      /  \\ \midrule
			$\mathrm{RoBERTa}_\mathrm{Large}$ &  8.1 &      17.8$\pm$4.0 &     8.3$\pm$3.1&   {\bf 8.0$\pm$3.1} &       / &     /  \\ 
			\bottomrule
		\end{tabular}
	}
\end{table*}

\begin{figure}
	\centering
	\includegraphics[width=.9\linewidth,trim={4cm .5cm 4cm 2cm},clip]{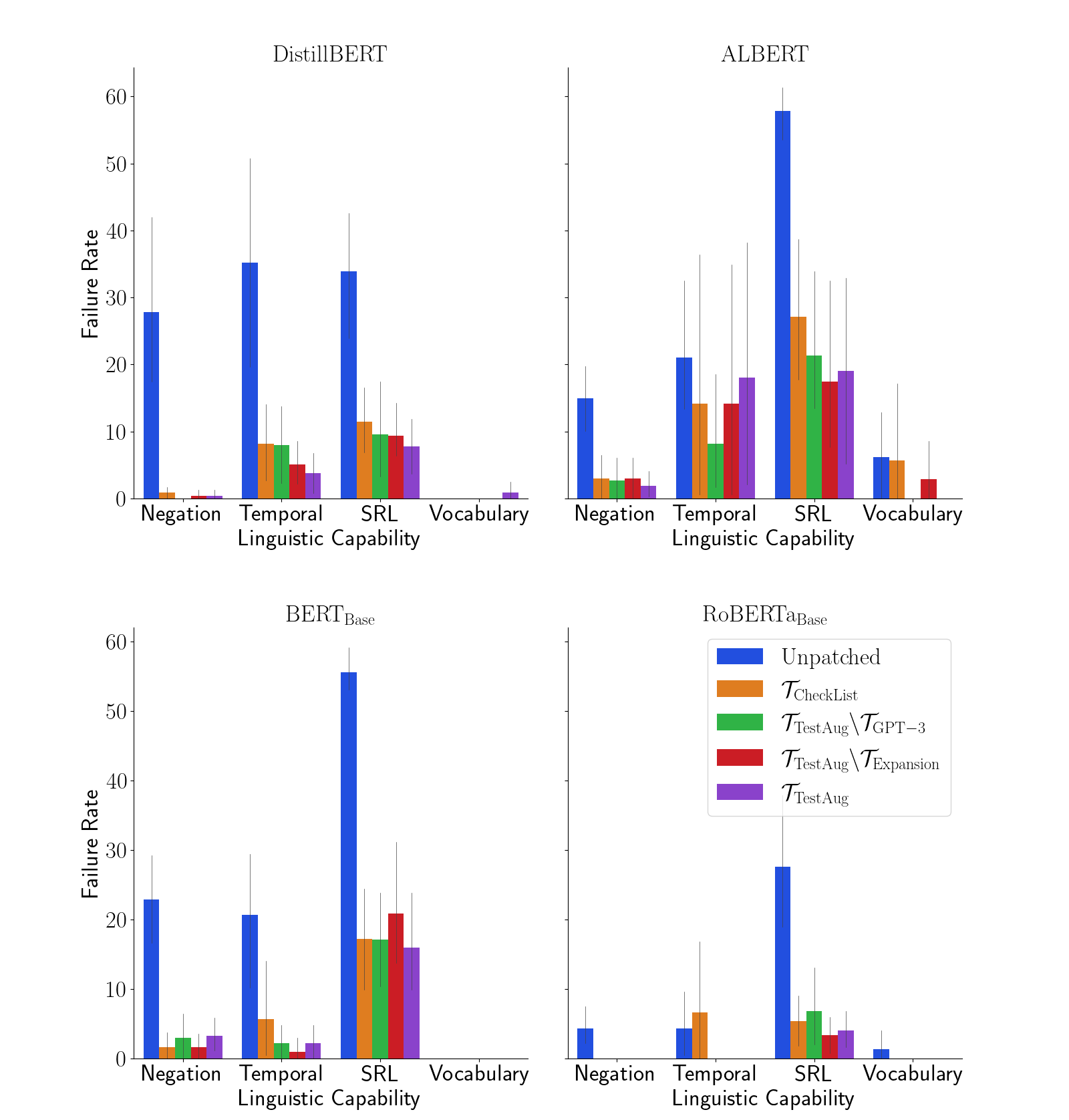}
	\caption{Capability-wise error rates of sentiment classification.}\label{fig:sentiment-cap}
\end{figure}
\subsection{Evaluating TestAug's Ability for Bug Detection}\label{sec:eval-bug}
\subsubsection{Metric}\label{sec:eval-method-bug}
\noindent \textbf{Patched Failure Rate}. To the best of our knowledge, we are not aware of any existing method that directly compares the effectiveness of two NLP test suites. 
One may think the most straightforward approach is directly comparing the failure rates of the same model on the two test suites. 
Despite the simplicity, we argue that these two failure rates are in fact \emph{incomparable}: the effectiveness of a test suite is defined by how many bugs it can find~\cite{kochhar2015code}; as a result, it is unclear whether a test suite with a lower failure rate but more error cases has a better performance. 


To compare the effectiveness of test suites $\fT_1, \fT_2, \cdots, \fT_N$, we propose to create new training and testing sets as follows.
\[
\fT_\mathrm{Test} \leftarrow \texttt{Sample}(\bigcup_{n=1}^N \fT_n),\quad
\fT_\mathrm{Train}^{(n)} \leftarrow \fT_n - \fT_\mathrm{Test}
\]

Then we use the training sets $\fT_\mathrm{Train}^{(n)}$ to patch (or fine-tune) the model $\fM$ and evaluate the patched (or fine-tuned) model $\hat{\fM}_n$ on the test set $\fT_\mathrm{Test}$.
We compare a model $\hat{\fM}_n$'s \textbf{patched failure rate} on different training sets; a lower rate thus indicates a stronger capability in finding bugs.
\[
\mathrm{FR}_\mathrm{Patched}^{(n)} = \sum_{i=1}^{\vert \fT_\mathrm{Test}\vert} 1(\hat{\fM}_n(x_i) \neq y_i)
\]

Additionally, we note that our evaluation method involves random partitions of $\fT$ and fine-tuning of the target models; the results of both depend on the choice of random seeds.
We, therefore, repeated our experiments with 5 different random seeds when partitioning $\fT$\footnote{These seeds $\{11, 14, 25, 42, 74\}$ are also randomly generated integers.};
we fixed the model fine-tuning seed to 42 to prevent evaluation results from being affected by the randomness of training.
\subsubsection{Analysis}
In Table~\ref{tab:aggregate}, Figure~\ref{fig:sentiment-cap} and Figure~\ref{fig:nli-qqp-cap} of the Appendix, we report the failure rates of TestAug and compare it with existing work~\cite{Ribeiro2020BeyondAB,Tarunesh2021LoNLIAE}. 
The three tasks we study contain 12 capabilities.

First, in Table~\ref{tab:aggregate}, we report the average failure rates across all capabilities for each task, compared with existing work. 
We use $\fT_\mathrm{TestAug}$ to represent the test suite by TestAug, and $\fT_\mathrm{CheckList}$ to represent the suite by existing work~\cite{Ribeiro2020BeyondAB,Tarunesh2021LoNLIAE}. 
We use the evaluation methodology in Section~\ref{sec:eval-method-bug} to merge $\fT_\mathrm{TestAug}$ and $\fT_\mathrm{CheckList}$ for comparing their failure rates. 
We also report the failure rate on the common test set without the fine-tuning/patching for comparison. 
In addition, we ablate study $\fT_\mathrm{TestAug}$'s performance by removing the cases directly from GPT-3 (i.e., $\fT_\mathrm{GPT-3}$) and the expansion (i.e., $\fT_\mathrm{Expansion}$), using only the resulting subset for patching. 
We can observe that the resulting failure rate of $\fT_\mathrm{TestAug}$ is consistently lower than $\fT_\mathrm{CheckList}$, indicating that $\fT_\mathrm{TestAug}$ can find more bugs.
We also find $\fT_\mathrm{TestAug}\backslash \fT_\mathrm{Expansion}$ to outperform $\fT_\mathrm{TestAug}$ in 4 cases, whereas $\fT_\mathrm{TestAug}\backslash \fT_\mathrm{GPT-3}$ outperforms the latter in only 1 case, this result indicates that GPT-3 generated cases are more important than the expanded cases, whereas the expanded cases are not always helpful. 
Next, we plot the capability-level failure rates for three tasks (Figure~\ref{fig:sentiment-cap}, Figure~\ref{fig:nli-qqp-cap} in Appendix)
From Figure~\ref{fig:sentiment-cap} and Figure~\ref{fig:nli-qqp-cap},
we can observe that $\fT_\mathrm{TestAug}$ does not consistently outperform $\fT_\mathrm{CheckList}$ when looking at each linguistic capability; for example, when evaluating sentiment classification task (Figure~\ref{fig:sentiment-cap}), the ALBERT's temporal capability gives us a higher failure rate of $\fT_\text{TestAug}$ after patching, indicating our augmented test suite has a weaker bug detection ability; a similar phenomenon could be found when testing vocabulary capability of $\mathrm{BERT}_\mathrm{Base}$ in paraphrase detection task and syntactic and presupposition capability of $\mathrm{RoBERTa}_\mathrm{Large}$ in natural language inference task.
This shows that, despite the overall ability to find more bugs (Table \ref{tab:aggregate}), the additional test cases from GPT-3 do not uniformly contribute to the improvement of each specific linguistic capability.
\subsection{Evaluating the Diversity of TestAug Results}\label{sec:eval-diverse}
\subsubsection{Metric}
As existing approaches rely on manually created templates, they have low linguistic variations.
Such issues can be alleviated with the help of GPT-3. 
In this section, we evaluate the linguistic diversity of the generated test cases.  
We use two metrics to evaluate the diversity: first, we leverage the Self-BLUE score to evaluate the diversity of an entire test suite; second, to measure the test case-level diversity, we introduce a new metric, i.e., the average number of unique dependency paths in each test case.

\noindent\textbf{Self-BLEU}. Self-BLEU is an extension of the regular BLEU that evaluates the diversity of generated texts~\cite{Zhu2018TexygenAB}. Given a list of texts $\hat{\fY}=\{\hat{Y}_1, \hat{Y}_2, \cdots, \hat{Y}_N\}$, Self-BLEU is the average BLEU score between every single sentence and all other sentences,
\begin{equation}
	\text{Self-BLEU}(\hat{\fY})=\frac{1}{N} \sum_{i=1}^N\text{BLEU}(\{\hat{Y}_i\}, \hat{\fY}_{\neq i})
\end{equation}
When $k$ is fixed, a lower Self-BLEU score indicates a higher diversity of the sentence.

\noindent\textbf{Number of Unique Dependency Paths}. We propose to use the number of unique dependency paths to measure the diversity at the test case level. 
Each path is a path of POS tags connected by edges in the dependency tree~\cite{Jurafsky2000SpeechAL}. For example, for the sentence "\emph{I (NOUN) love (VERB) chicken (NOUN)}", the unique paths are {NOUN\textrightarrow VERB, VERB\textrightarrow NOUN, NOUN\textrightarrow VERB \textrightarrow NOUN}. 
\subsubsection{Analysis}
We compare the Self-BLEU and the number of unique dependency paths between TestAug and CheckList in Table~\ref{tab:diversity},
where we control the number of test cases under each test\footnote{We sampled 100 unique sentences per test with a fixed seed of 42 for both test suites. This gave us 1100, 1700, and 1200 sentences in sentiment classification, paraphrase detection, and natural language inference task, respectively.}. 
From Table~\ref{tab:diversity}, we can observe that the linguistic diversity of the test suites $\fT_\mathrm{GPT-3}$ show \textbf{substantial} improvement over the template-based counterparts $\fT_\mathrm{CheckList}$: the Self-BLEU4 score has a decrease at least 5.3\% (the natural language inference task) and the number of unique dependency paths is of at least 1.81 times compared to the original test suite (the natural language inference task).
\begin{table}[!htb]
	\caption{TestAug saves manual efforts in generating new test cases and expands the set of available templates.}
	\label{tab:saving}
	\resizebox{\linewidth}{!}{
		\begin{tabular}{ccccc}
			\toprule
			\makecell[c]{\#Uniuqe Seed\\ Sentences} &  \makecell[c]{\#Unique New\\ Sentences} &  
			\makecell[c]{\#Seed\\ Templates} &  
			\makecell[c]{\#New\\ Templates} \\ 
			\midrule
			\multicolumn{4}{c}{Sentiment Classification}\\
			\midrule
			292 &                   3275 (11.2\texttimes) &               29 &            1441 (49.7\texttimes) \\ 
			\midrule
			\multicolumn{4}{c}{Paraphrase Detection}\\
			\midrule
			124 &                   6427 (50.4\texttimes) &               17 &            1307 (76.9\texttimes)\\ 
			
			\midrule
			\multicolumn{4}{c}{Natural Language Inference}\\
			\midrule
			150 &                   4976 (33.2\texttimes) &               50 &               \makecell[c]{/} \\ 
			
			\bottomrule
	\end{tabular}}
\end{table}

\begin{table}[H]
	\centering
	\caption{Linguistic diversity of test suites.}\label{tab:diversity}
	\resizebox{\columnwidth}{!}{
		\begin{tabular}{lcc}
			\toprule
			{} &  Self-BLEU4 (\textdownarrow) &  \makecell{Number of Unique\\ Dependency Paths} (\textuparrow) \\
			\midrule
			\multicolumn{3}{c}{Sentiment Analysis}\\
			\midrule
			
			$\fT_\mathrm{TestAug}$    &       0.634 &  480 \\ 
			$\fT_\mathrm{CheckList}$ &       0.853 &   84 \\

			\midrule
			\multicolumn{3}{c}{Paraphrase Detection}\\
			\midrule
			
			$\fT_\mathrm{TestAug}$    &       0.627 &  418 \\ 
			$\fT_\mathrm{CheckList}$ &       0.775 &  117 \\ 
			
			\midrule
			\multicolumn{3}{c}{Natural Language Inference}\\
			\midrule
			
			$\fT_\mathrm{TestAug}$    &       0.430 &  210 \\
			$\fT_\mathrm{CheckList}$ &       0.454 &  116 \\ 
			
			
			\bottomrule
		\end{tabular}
	}
\end{table}

\subsection{Evaluating the Validity of TestAug Results}
\label{sec:eval-val}
In this section, we evaluate the effectiveness of our pipeline of filtering incorrect test cases (i.e., Section~\ref{sec:augment}). 
As is shown in Table \ref{tab:acc} of the Appendix, the two-phase approach successfully filters the invalid cases: the classification accuracy is \emph{consistently} higher than the validity threshold (90\% for sentiment classification and paraphrase detection and 80\% for natural language inference)\footnote{Additional capabilities in NLI tasks are below 80\% valid threshold. 
	Automatic filtering these cases out with trained classifiers are left as future work.}.
At the same time, the Cohen's $\kappa$ on test set annotation indicates \emph{substantial} agreement ~\cite{McHugh2012InterraterRT}.

\begin{table}
	\caption{Annotation Cohen's $\kappa$ and agreement rate.}
	\label{tab:kappa}
	\resizebox{\columnwidth}{!}{
		\begin{tabular}{lcc}
			\toprule
			{} & $\frac{\mathrm{Agreement}}{\mathrm{Total}}$ & Cohen's $\kappa$ \\ 
			\midrule
			Sentiment Analysis &                $\frac{ 438 }{ 461 }=95.0\%$ &            0.741 \\  \midrule
			Paraphrase Detection &                $\frac{ 365 }{ 401 }=91.0\%$ &            0.812 \\ \midrule
			Natural Language Inference &                $\frac{ 151 }{ 156 }=96.8\%$ &            0.746 \\
			\bottomrule
	\end{tabular}}
\end{table}

\subsection{Evaluation of the Manual Efforts Saved by TestAug}
\label{eval:efforts}
Finally, we quantitatively evaluate the manual efforts saved by TestAug compared to CheckList; the results are reported in Table~\ref{tab:saving}. 
When querying GPT-3 with a few hundred sentences per task, we obtained a new set of valid test cases 11.2 to 50.4 times in size. 
The template expansion in turn increased the number of available templates to at least 49.7 times, reducing manual efforts by at least 98.0\% (Table \ref{tab:saving})\footnote{New templates do not cost additional manual efforts when considering all templates as a whole, leading to at least $\frac{1441}{1441+29}\times 100\%=98.0\%$ saving.
	With the help of our automatic filtering pipeline, both sentence and template counts could be further increased with additional queries.}.
This result thus shows that TestAug can largely save manual efforts in creating the test suites. 

\section{Discussion, Conclusions, and Future Work}
This paper introduces a novel framework TestAug for capability-based NLP testing. 
Addressing current system's heavy dependence on manual creation of templates, our framework can save at least 98.0\% of the manual annotation effort with GPT-3; meanwhile, the test suites generated with our framework reveal more bugs than existing systems and show better diversity. 
Our framework guarantees the correctness of the generated cases by removing the invalid output from GPT-3. 

The main limitation of TestAug is that GPT-3 fails to generate highly structured test cases, such as cases for extractive question answering. 
It also struggles to generate cases that require logic or math reasoning. 
We leave the exploration of these cases for future work. 


\clearpage
\section*{Ethical Considerations}

\subsection*{Annotator Rights}
Two of the authors (one male and one female; both identified themselves as Asians) annotated the data following annotation guidelines; the guidelines are discussed and finalized after thorough discussions (the violations of these guidelines are discussed in Section \ref{sec:augment}).
We acknowledge the annotators' efforts with shared authorship.

\subsection*{Intended Uses}
TestAug's intended use is as a tool to augment template-based test suites with newly generated test cases from GPT-3; two sets of test cases are then used altogether to evaluate an NLP models' linguistic capabilities; we believe this application of existing datasets are consistent with their intended uses.
We showed the effectiveness of this system in Section \ref{sec:experiment}.
We hope the adoption of TestAug into the NLP model development could make newly built NLP models more linguistically capable.
Meanwhile, the TestAug includes GPT-3 as a component, we urge users of our system to follow the OpenAI's usage guidelines \footnote{\url{https://beta.openai.com/docs/usage-guidelines}}.

\subsection*{Potential Misuse}
TestAug might be misused to overestimate the models' linguistic capabilities.
Specifically, even though failures on the test suites show models' shortcomings in a given linguistic capability, the absence of failures does \emph{not} mean the models being tested are free from bugs; it is likely that test suites are not yet capable enough to reveal the model's bugs.
We, therefore, call for a judicious interpretation of an NLP model's performance based on TestAug test suites.
Moreover, we believe NLP testing is an iterative process; it might take multiple iterations of applying TestAug to reveal the model's issues in linguistic capabilities.

\bibliography{citation}

\begin{thebibliography}{38}
\expandafter\ifx\csname natexlab\endcsname\relax\def\natexlab#1{#1}\fi

\bibitem[{Bai et~al.(2021)Bai, Liang, Zhang, Li, Bai, and Wang}]{Bai2021WhyAM}
Bing Bai, Jian Liang, Guan Zhang, Hao Li, Kun Bai, and Fei Wang. 2021.
\newblock Why attentions may not be interpretable?
\newblock \emph{Proceedings of the 27th ACM SIGKDD Conference on Knowledge
  Discovery \& Data Mining}.

\bibitem[{Bowman et~al.(2015)Bowman, Angeli, Potts, and
  Manning}]{Bowman2015ALA}
Samuel~R. Bowman, Gabor Angeli, Christopher Potts, and Christopher~D. Manning.
  2015.
\newblock A large annotated corpus for learning natural language inference.
\newblock In \emph{EMNLP}.

\bibitem[{Bowman and Dahl(2021)}]{bowman2021will}
Samuel~R Bowman and George~E Dahl. 2021.
\newblock \href {https://aclanthology.org/2021.naacl-main.385.pdf} {What will
  it take to fix benchmarking in natural language understanding?}
\newblock \emph{The 2020 Conference of the North American Chapter of the
  Association for Computational Linguistics - Human Language Technologies
  (NAACL-HLT2020)}.

\bibitem[{Brown et~al.(2020)Brown, Mann, Ryder, Subbiah, Kaplan, Dhariwal,
  Neelakantan, Shyam, Sastry, Askell et~al.}]{brown2020language}
Tom~B Brown, Benjamin Mann, Nick Ryder, Melanie Subbiah, Jared Kaplan, Prafulla
  Dhariwal, Arvind Neelakantan, Pranav Shyam, Girish Sastry, Amanda Askell,
  et~al. 2020.
\newblock \href {https://arxiv.org/abs/2005.14165} {Language models are
  few-shot learners}.
\newblock \emph{arXiv preprint arXiv:2005.14165}.

\bibitem[{Gao et~al.(2021)Gao, Fisch, and Chen}]{gao-etal-2021-making}
Tianyu Gao, Adam Fisch, and Danqi Chen. 2021.
\newblock \href {https://doi.org/10.18653/v1/2021.acl-long.295} {Making
  pre-trained language models better few-shot learners}.
\newblock In \emph{Proceedings of the 59th Annual Meeting of the Association
  for Computational Linguistics and the 11th International Joint Conference on
  Natural Language Processing (Volume 1: Long Papers)}, pages 3816--3830,
  Online. Association for Computational Linguistics.

\bibitem[{Geva et~al.(2019)Geva, Goldberg, and Berant}]{Geva2019AreWM}
Mor Geva, Yoav Goldberg, and Jonathan Berant. 2019.
\newblock Are we modeling the task or the annotator? an investigation of
  annotator bias in natural language understanding datasets.
\newblock \emph{ArXiv}, abs/1908.07898.

\bibitem[{Glockner et~al.(2018)Glockner, Shwartz, and
  Goldberg}]{Glockner2018BreakingNS}
Max Glockner, Vered Shwartz, and Yoav Goldberg. 2018.
\newblock Breaking nli systems with sentences that require simple lexical
  inferences.
\newblock In \emph{ACL}.

\bibitem[{Gururangan et~al.(2018)Gururangan, Swayamdipta, Levy, Schwartz,
  Bowman, and Smith}]{Gururangan2018AnnotationAI}
Suchin Gururangan, Swabha Swayamdipta, Omer Levy, Roy Schwartz, Samuel~R.
  Bowman, and Noah~A. Smith. 2018.
\newblock Annotation artifacts in natural language inference data.
\newblock In \emph{NAACL}.

\bibitem[{Jeretic et~al.(2020)Jeretic, Warstadt, Bhooshan, and
  Williams}]{jeretic-etal-2020-natural}
Paloma Jeretic, Alex Warstadt, Suvrat Bhooshan, and Adina Williams. 2020.
\newblock \href {https://doi.org/10.18653/v1/2020.acl-main.768} {Are natural
  language inference models {IMPPRESsive}? {L}earning {IMPlicature} and
  {PRESupposition}}.
\newblock In \emph{Proceedings of the 58th Annual Meeting of the Association
  for Computational Linguistics}, pages 8690--8705, Online. Association for
  Computational Linguistics.

\bibitem[{Joshi et~al.(2020{\natexlab{a}})Joshi, Aditya, Sathe, and
  Choudhury}]{joshi-etal-2020-taxinli}
Pratik Joshi, Somak Aditya, Aalok Sathe, and Monojit Choudhury.
  2020{\natexlab{a}}.
\newblock \href {https://doi.org/10.18653/v1/2020.conll-1.4} {{T}axi{NLI}:
  Taking a ride up the {NLU} hill}.
\newblock In \emph{Proceedings of the 24th Conference on Computational Natural
  Language Learning}, pages 41--55, Online. Association for Computational
  Linguistics.

\bibitem[{Joshi et~al.(2020{\natexlab{b}})Joshi, Aditya, Sathe, and
  Choudhury}]{Joshi2020TaxiNLITA}
Pratik~M. Joshi, Somak Aditya, Aalok Sathe, and Monojit Choudhury.
  2020{\natexlab{b}}.
\newblock Taxinli: Taking a ride up the nlu hill.
\newblock In \emph{CONLL}.

\bibitem[{Jurafsky and Martin(2000)}]{Jurafsky2000SpeechAL}
Dan Jurafsky and James~H. Martin. 2000.
\newblock Speech and language processing.

\bibitem[{Kirk et~al.(2021)Kirk, Vidgen, R{\"o}ttger, Thrush, and
  Hale}]{kirk2021hatemoji}
Hannah~Rose Kirk, Bertram Vidgen, Paul R{\"o}ttger, Tristan Thrush, and Scott~A
  Hale. 2021.
\newblock \href {https://arxiv.org/pdf/2108.05921} {Hatemoji: A test suite and
  adversarially-generated dataset for benchmarking and detecting emoji-based
  hate}.
\newblock \emph{arXiv preprint arXiv:2108.05921}.

\bibitem[{Kochhar et~al.(2015)Kochhar, Thung, and Lo}]{kochhar2015code}
Pavneet~Singh Kochhar, Ferdian Thung, and David Lo. 2015.
\newblock Code coverage and test suite effectiveness: Empirical study with real
  bugs in large systems.
\newblock In \emph{2015 IEEE 22nd international conference on software
  analysis, evolution, and reengineering (SANER)}, pages 560--564. IEEE.

\bibitem[{Lewis et~al.(2020)Lewis, Liu, Goyal, Ghazvininejad, Mohamed, Levy,
  Stoyanov, and Zettlemoyer}]{lewis-etal-2020-bart}
Mike Lewis, Yinhan Liu, Naman Goyal, Marjan Ghazvininejad, Abdelrahman Mohamed,
  Omer Levy, Veselin Stoyanov, and Luke Zettlemoyer. 2020.
\newblock \href {https://doi.org/10.18653/v1/2020.acl-main.703} {{BART}:
  Denoising sequence-to-sequence pre-training for natural language generation,
  translation, and comprehension}.
\newblock In \emph{Proceedings of the 58th Annual Meeting of the Association
  for Computational Linguistics}, pages 7871--7880, Online. Association for
  Computational Linguistics.

\bibitem[{Li and Liang(2021)}]{li-liang-2021-prefix}
Xiang~Lisa Li and Percy Liang. 2021.
\newblock \href {https://doi.org/10.18653/v1/2021.acl-long.353} {Prefix-tuning:
  Optimizing continuous prompts for generation}.
\newblock In \emph{Proceedings of the 59th Annual Meeting of the Association
  for Computational Linguistics and the 11th International Joint Conference on
  Natural Language Processing (Volume 1: Long Papers)}, pages 4582--4597,
  Online. Association for Computational Linguistics.

\bibitem[{Liu et~al.(2022)Liu, Swayamdipta, Smith, and Choi}]{Liu2022WANLIWA}
Alisa Liu, Swabha Swayamdipta, Noah~A. Smith, and Yejin Choi. 2022.
\newblock Wanli: Worker and ai collaboration for natural language inference
  dataset creation.

\bibitem[{Liu et~al.(2021)Liu, Yuan, Fu, Jiang, Hayashi, and
  Neubig}]{Liu2021PretrainPA}
Pengfei Liu, Weizhe Yuan, Jinlan Fu, Zhengbao Jiang, Hiroaki Hayashi, and
  Graham Neubig. 2021.
\newblock Pre-train, prompt, and predict: A systematic survey of prompting
  methods in natural language processing.
\newblock \emph{ArXiv}, abs/2107.13586.

\bibitem[{McHugh(2012)}]{McHugh2012InterraterRT}
M.~L. McHugh. 2012.
\newblock Interrater reliability: the kappa statistic.
\newblock \emph{Biochemia Medica}, 22:276 -- 282.

\bibitem[{Mishra et~al.(2021)Mishra, Khashabi, Baral, and
  Hajishirzi}]{mishra2021cross}
Swaroop Mishra, Daniel Khashabi, Chitta Baral, and Hannaneh Hajishirzi. 2021.
\newblock \href {https://arxiv.org/pdf/2104.08773.pdf} {Cross-task
  generalization via natural language crowdsourcing instructions}.
\newblock \emph{arXiv preprint arXiv:2104.08773}.

\bibitem[{Poliak et~al.(2018)Poliak, Haldar, Rudinger, Hu, Pavlick, White, and
  Van~Durme}]{poliak-etal-2018-collecting}
Adam Poliak, Aparajita Haldar, Rachel Rudinger, J.~Edward Hu, Ellie Pavlick,
  Aaron~Steven White, and Benjamin Van~Durme. 2018.
\newblock \href {https://doi.org/10.18653/v1/D18-1007} {Collecting diverse
  natural language inference problems for sentence representation evaluation}.
\newblock In \emph{Proceedings of the 2018 Conference on Empirical Methods in
  Natural Language Processing}, pages 67--81, Brussels, Belgium. Association
  for Computational Linguistics.

\bibitem[{Reif et~al.(2021)Reif, Ippolito, Yuan, Coenen, Callison-Burch, and
  Wei}]{Reif2021ARF}
Emily Reif, Daphne Ippolito, Ann Yuan, Andy Coenen, Chris Callison-Burch, and
  Jason Wei. 2021.
\newblock A recipe for arbitrary text style transfer with large language
  models.
\newblock \emph{ArXiv}, abs/2109.03910.

\bibitem[{Reynolds and McDonell(2021{\natexlab{a}})}]{reynolds2021prompt}
Laria Reynolds and Kyle McDonell. 2021{\natexlab{a}}.
\newblock \href
  {https://dl.acm.org/doi/pdf/10.1145/3411763.3451760?casa_token=gNN8C9ceCTwAAAAA:AujhM5_vBvVBYDNsjX2rNsiToPgbxP5ge8S7pXRIlEG1RzH2ljSRUD__1XGhZjZb7U5C8dFM4sy7}
  {Prompt programming for large language models: Beyond the few-shot paradigm}.
\newblock In \emph{Extended Abstracts of the 2021 CHI Conference on Human
  Factors in Computing Systems}, pages 1--7.

\bibitem[{Reynolds and McDonell(2021{\natexlab{b}})}]{Reynolds2021PromptPF}
Laria Reynolds and Kyle McDonell. 2021{\natexlab{b}}.
\newblock Prompt programming for large language models: Beyond the few-shot
  paradigm.
\newblock \emph{Extended Abstracts of the 2021 CHI Conference on Human Factors
  in Computing Systems}.

\bibitem[{Ribeiro et~al.(2020)Ribeiro, Wu, Guestrin, and
  Singh}]{Ribeiro2020BeyondAB}
Marco~Tulio Ribeiro, Tongshuang Wu, Carlos Guestrin, and Sameer Singh. 2020.
\newblock \href {https://doi.org/10.18653/v1/2020.acl-main.442} {Beyond
  accuracy: Behavioral testing of {NLP} models with {C}heck{L}ist}.
\newblock In \emph{Proceedings of the 58th Annual Meeting of the Association
  for Computational Linguistics}, pages 4902--4912, Online. Association for
  Computational Linguistics.

\bibitem[{Richardson et~al.(2020)Richardson, Hu, Moss, and
  Sabharwal}]{richardson2020probing}
Kyle Richardson, Hai Hu, Lawrence Moss, and Ashish Sabharwal. 2020.
\newblock \href {https://ojs.aaai.org/index.php/AAAI/article/view/6397}
  {Probing natural language inference models through semantic fragments}.
\newblock In \emph{Proceedings of the AAAI Conference on Artificial
  Intelligence}, volume~34, pages 8713--8721.

\bibitem[{R{\"o}ttger et~al.(2021)R{\"o}ttger, Vidgen, Nguyen, Waseem,
  Margetts, and Pierrehumbert}]{rottger-etal-2021-hatecheck}
Paul R{\"o}ttger, Bertie Vidgen, Dong Nguyen, Zeerak Waseem, Helen Margetts,
  and Janet Pierrehumbert. 2021.
\newblock \href {https://doi.org/10.18653/v1/2021.acl-long.4} {{H}ate{C}heck:
  Functional tests for hate speech detection models}.
\newblock In \emph{Proceedings of the 59th Annual Meeting of the Association
  for Computational Linguistics and the 11th International Joint Conference on
  Natural Language Processing (Volume 1: Long Papers)}, pages 41--58, Online.
  Association for Computational Linguistics.

\bibitem[{Salvatore et~al.(2019)Salvatore, Finger, and
  Hirata~Jr}]{salvatore-etal-2019-logical}
Felipe Salvatore, Marcelo Finger, and Roberto Hirata~Jr. 2019.
\newblock \href {https://doi.org/10.18653/v1/D19-6103} {A logical-based corpus
  for cross-lingual evaluation}.
\newblock In \emph{Proceedings of the 2nd Workshop on Deep Learning Approaches
  for Low-Resource NLP (DeepLo 2019)}, pages 22--30, Hong Kong, China.
  Association for Computational Linguistics.

\bibitem[{Schick and Sch{\"u}tze(2021)}]{Schick2021GeneratingDW}
Timo Schick and Hinrich Sch{\"u}tze. 2021.
\newblock Generating datasets with pretrained language models.
\newblock In \emph{EMNLP}.

\bibitem[{Shin et~al.(2020)Shin, Razeghi, Logan~IV, Wallace, and
  Singh}]{shin-etal-2020-autoprompt}
Taylor Shin, Yasaman Razeghi, Robert~L. Logan~IV, Eric Wallace, and Sameer
  Singh. 2020.
\newblock \href {https://doi.org/10.18653/v1/2020.emnlp-main.346}
  {{A}uto{P}rompt: {E}liciting {K}nowledge from {L}anguage {M}odels with
  {A}utomatically {G}enerated {P}rompts}.
\newblock In \emph{Proceedings of the 2020 Conference on Empirical Methods in
  Natural Language Processing (EMNLP)}, pages 4222--4235, Online. Association
  for Computational Linguistics.

\bibitem[{Socher et~al.(2013)Socher, Perelygin, Wu, Chuang, Manning, Ng, and
  Potts}]{Socher2013RecursiveDM}
Richard Socher, Alex Perelygin, Jean Wu, Jason Chuang, Christopher~D. Manning,
  Andrew Ng, and Christopher Potts. 2013.
\newblock \href {https://aclanthology.org/D13-1170} {Recursive deep models for
  semantic compositionality over a sentiment treebank}.
\newblock In \emph{Proceedings of the 2013 Conference on Empirical Methods in
  Natural Language Processing}, pages 1631--1642, Seattle, Washington, USA.
  Association for Computational Linguistics.

\bibitem[{Tarunesh et~al.(2021)Tarunesh, Aditya, and
  Choudhury}]{Tarunesh2021LoNLIAE}
Ishan Tarunesh, Somak Aditya, and Monojit Choudhury. 2021.
\newblock \href {https://arxiv.org/pdf/2112.02333.pdf} {Lonli: An extensible
  framework for testing diverse logical reasoning capabilities for nli}.
\newblock In \emph{Thirty-Sixth AAAI Conference on Artificial Intelligence
  (AAAI-22)}, volume abs/2112.02333.

\bibitem[{van Aken et~al.(2021)van Aken, Herrmann, and L{\"o}ser}]{van2021you}
Betty van Aken, Sebastian Herrmann, and Alexander L{\"o}ser. 2021.
\newblock \href {https://arxiv.org/pdf/2111.15512.pdf} {What do you see in this
  patient? behavioral testing of clinical nlp models}.
\newblock \emph{NeurIPS 2021 Research2Clinics Workshop, Bridging the Gap: From
  Machine Learning Research to Clinical Practice}.

\bibitem[{Wang et~al.(2021)Wang, Xu, Guzm{\'a}n, El-Kishky, Rubinstein, and
  Cohn}]{wang-etal-2021-easy}
Jun Wang, Chang Xu, Francisco Guzm{\'a}n, Ahmed El-Kishky, Benjamin Rubinstein,
  and Trevor Cohn. 2021.
\newblock \href {https://doi.org/10.18653/v1/2021.findings-acl.415} {As easy as
  1, 2, 3: Behavioural testing of {NMT} systems for numerical translation}.
\newblock In \emph{Findings of the Association for Computational Linguistics:
  ACL-IJCNLP 2021}, pages 4711--4717, Online. Association for Computational
  Linguistics.

\bibitem[{West et~al.(2021)West, Bhagavatula, Hessel, Hwang, Jiang, Bras, Lu,
  Welleck, and Choi}]{West2021SymbolicKD}
Peter West, Chandrasekhar Bhagavatula, Jack Hessel, Jena~D. Hwang, Liwei Jiang,
  Ronan~Le Bras, Ximing Lu, Sean Welleck, and Yejin Choi. 2021.
\newblock Symbolic knowledge distillation: from general language models to
  commonsense models.
\newblock \emph{ArXiv}, abs/2110.07178.

\bibitem[{Yanaka et~al.(2019)Yanaka, Mineshima, Bekki, Inui, Sekine,
  Abzianidze, and Bos}]{yanaka-etal-2019-help}
Hitomi Yanaka, Koji Mineshima, Daisuke Bekki, Kentaro Inui, Satoshi Sekine,
  Lasha Abzianidze, and Johan Bos. 2019.
\newblock \href {https://doi.org/10.18653/v1/S19-1027} {{HELP}: A dataset for
  identifying shortcomings of neural models in monotonicity reasoning}.
\newblock In \emph{Proceedings of the Eighth Joint Conference on Lexical and
  Computational Semantics (*{SEM} 2019)}, pages 250--255, Minneapolis,
  Minnesota. Association for Computational Linguistics.

\bibitem[{Zhang et~al.(2019)Zhang, Harman, Ma, and Liu}]{Zhang2019MachineLT}
Jie~M. Zhang, Mark Harman, Lei Ma, and Yang Liu. 2019.
\newblock Machine learning testing: Survey, landscapes and horizons.
\newblock \emph{ArXiv}, abs/1906.10742.

\bibitem[{Zhu et~al.(2018)Zhu, Lu, Zheng, Guo, Zhang, Wang, and
  Yu}]{Zhu2018TexygenAB}
Yaoming Zhu, Sidi Lu, Lei Zheng, Jiaxian Guo, Weinan Zhang, Jun Wang, and Yong
  Yu. 2018.
\newblock Texygen: A benchmarking platform for text generation models.
\newblock \emph{The 41st International ACM SIGIR Conference on Research \&
  Development in Information Retrieval}.

\end{thebibliography}
\clearpage

\appendix
\section{Appendix}
\begin{figure*}
	\centering
	\begin{subfigure}[b]{\textwidth}
		\includegraphics[width=\linewidth,trim={.5cm 0cm .5cm 1.5cm},clip]{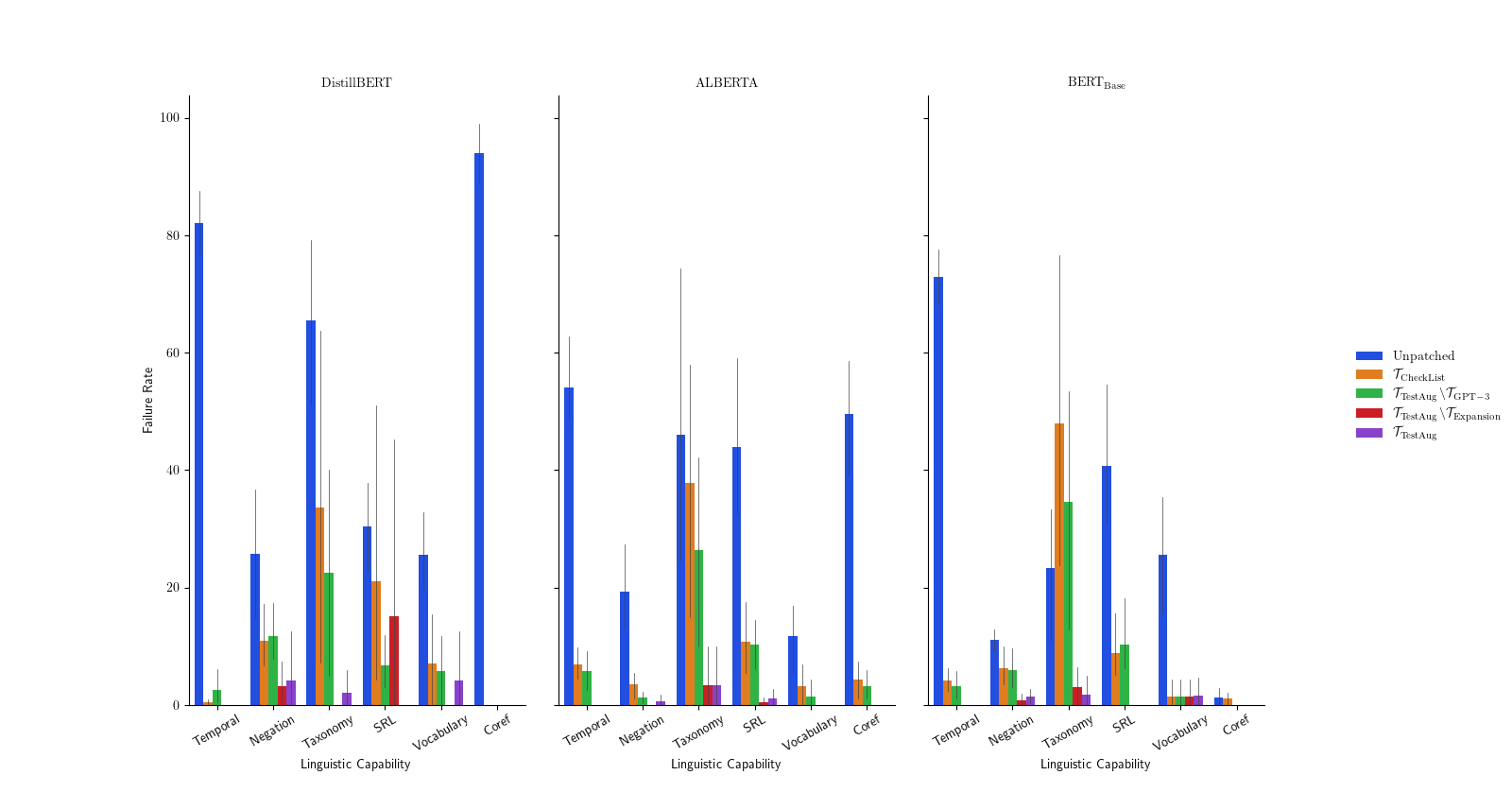}
		\subcaption{Paraphrase Detection}
	\end{subfigure}
	\begin{subfigure}[b]{\textwidth}
		\includegraphics[width=\linewidth,trim={.5cm 0cm .5cm 1.5cm},clip]{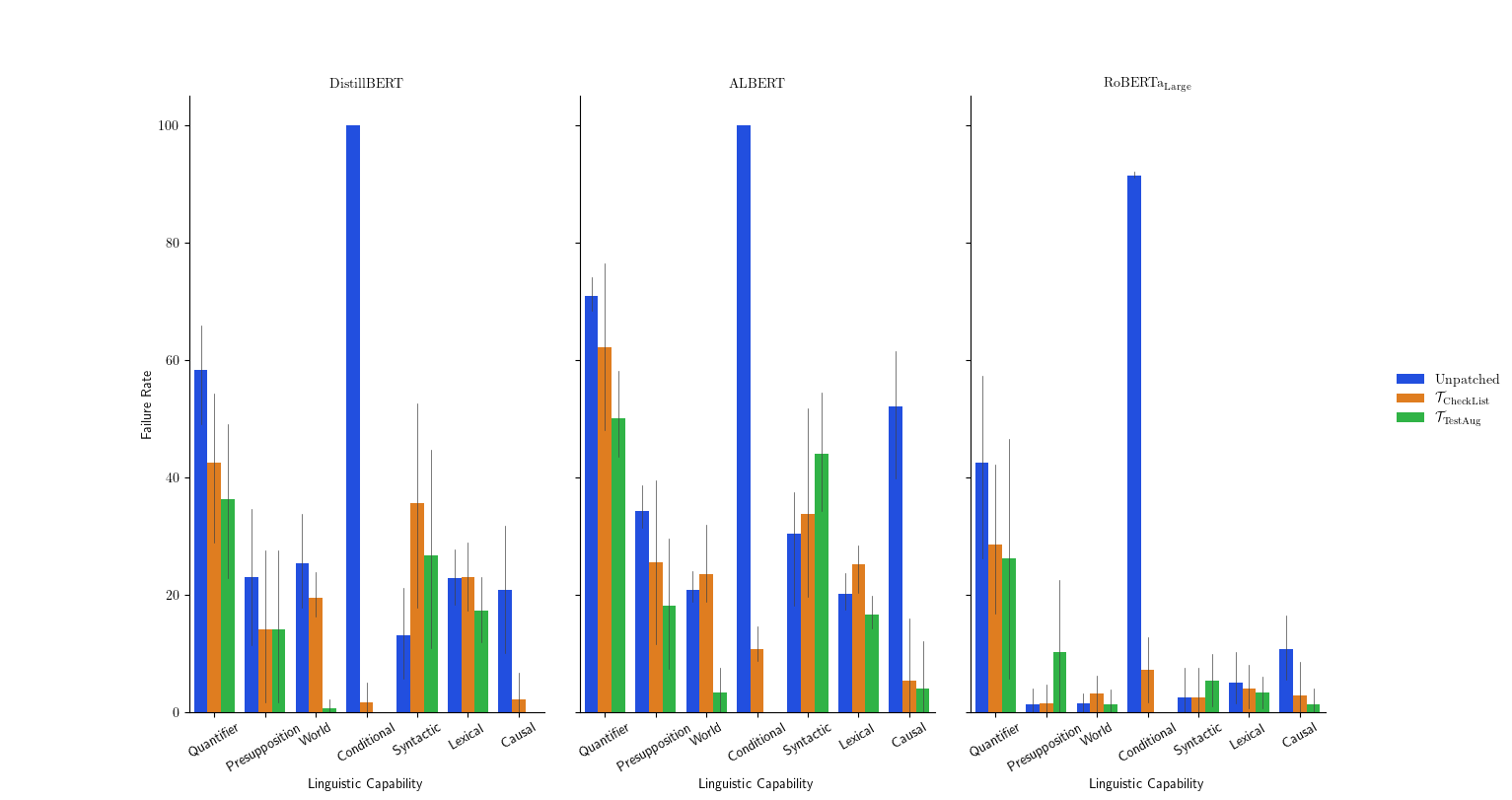}
		\subcaption{Natural Language Inference}
	\end{subfigure}
	\caption{Capability-wise error rates on paraphrase detection and natural language inference tasks.}
	\label{fig:nli-qqp-cap}
\end{figure*}

\begin{table}[H]
	\centering
	\caption{Prompt designs for paraphrase detection and natural language inference tasks.}
	\label{tab:nli-qqp-prompt}
	\resizebox{\columnwidth}{!}{
		\begin{tabular}{c}
			\midrule
			Paraphrase Detection\\
			\midrule
			\includegraphics[width=\columnwidth]{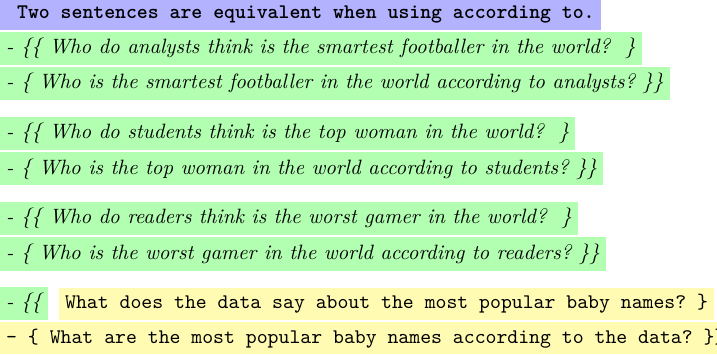}\\
			\midrule
			Natural Language Inference\\
			\midrule
			\includegraphics[width=\columnwidth]{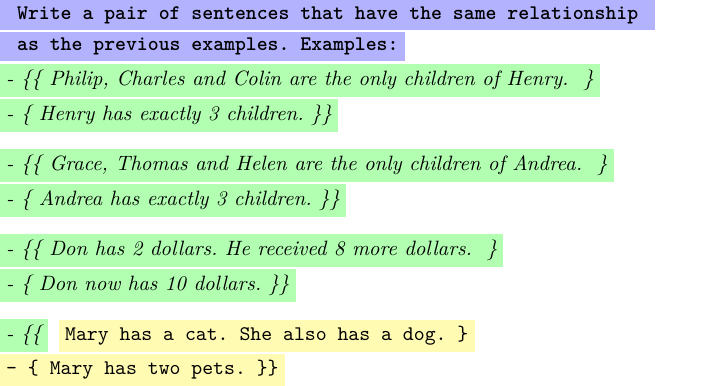}\\
			\bottomrule
	\end{tabular}}
\end{table}

\begin{table}[H]
	\centering
	\caption{Hyperperparamer choice for model fine-tuning}\label{tab:hparam}
	\begin{tabular}{lc}
		\toprule
		Hyperparameter& Value\\
		\midrule
		Learning rate& $5e-6$\\
		Batch size& 16\\
		Number of training epochs& 10\\
		Max. sequence length& 128\\
		Seed& 42\\
		\bottomrule
	\end{tabular}
\end{table}

\begin{figure*}
	\centering
	\includegraphics[width=\textwidth]{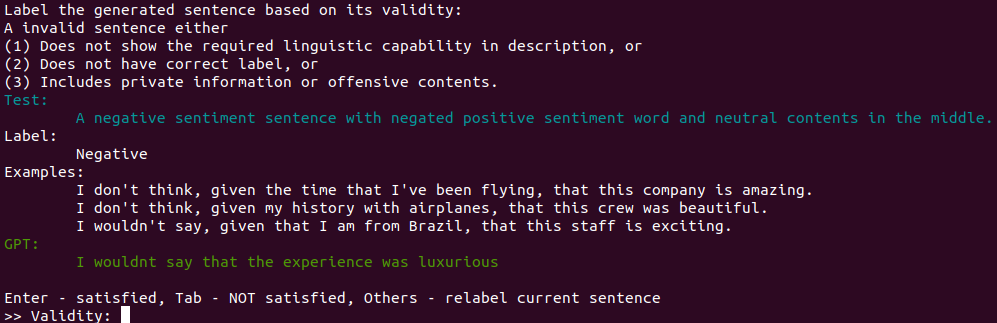}
	\caption{The command line interface for data annotation. 
		Annotators are given a test and three associated test cases from the template-based test suite; they are asked to the annotate the validity of the GPT-3-generated sentences.
		Annotators are reminded of the guidelines for filtering invalid samples when labeling each sentence (shown at the top of the interface).
		We communicated explicitly for the intended uses of the annotated datasets before the annotation.}\label{fig:ui}
\end{figure*}

\begin{table*}
	\caption{Linguistic capabilities that appeared in the experiments and their explanations with corresponding examples.
		The description of the tested linguistic capabilities follow the taxonomy and definition provided in previous work \cite{Ribeiro2020BeyondAB, Tarunesh2021LoNLIAE, Joshi2020TaxiNLITA}.}
	\label{tab:linguistic}
	\resizebox{\textwidth}{!}{
		\begin{tabular}{lll}
			\toprule
			Linguistic Capability & Explanation &                                                                                                                                                                                                                                                                                                                                                                            Template and Example \\ 
			\midrule
			\multicolumn{3}{c}{Sentiment Analysis}\\
			\midrule
			Negation &   Resolving "not"         &                                                                                 \begin{tabular}{@{}l@{}}\{neg\} \{pos\_verb\_present\} \{the\} \{air\_noun\}.\\I didn't admire that service.\end{tabular} \\ \midrule
			SRL &  Resolving subjects and objects            &                                                                \begin{tabular}{@{}l@{}}Do I think \{it\} \{be\} \{a:pos\_adj\} \{air\_noun\}? No\\Do I think that is an amazing aircraft? No\end{tabular} \\ \midrule
			Temporal &   Resolving temporal changes         & \begin{tabular}{@{}l@{}}I \{neg\_verb\_present\} this airline, \{change\} in the past I would \{pos\_verb\_present\} it.\\I regret this airline, although in the past I would appreciate it.\end{tabular} \\ \midrule
			Vocabulary &  Resolving word choice variations           &                                                                                                       \begin{tabular}{@{}l@{}}\{it\} \{air\_noun\} \{be\} \{neg\_adj\}.\\That food was ugly.\end{tabular} \\ 
			
			\midrule
			\multicolumn{3}{c}{Paraphrase Detection}\\
			\midrule
			Coref & Resolving male and female names             & \begin{tabular}{@{}l@{}}\begin{tabular}{@{}l@{}}S1: If \{female\} and \{male\} were alone, do you think he would reject her?\\S2: If \{female\} and \{male\} were alone, do you think she would reject him?\end{tabular}\\\begin{tabular}{@{}l@{}}S1: If Julie and Roy were alone, do you think he would reject her?\\S2: If Julie and Roy were alone, do you think she would reject him?\end{tabular}\end{tabular} \\ \midrule
			Negation &   Resolving "not"          &                                                                                   \begin{tabular}{@{}l@{}}\begin{tabular}{@{}l@{}}S1: What are things \{a:noun\} should worry about?\\S2: What are things \{a:noun\} should not worry about?\end{tabular}\\\begin{tabular}{@{}l@{}}S1: What are things a friend should worry about?\\S2: What are things a friend should not worry about?\end{tabular}\end{tabular} \\ \midrule
			SRL &   Resolving subjects and objects          &                                                                                                                     \begin{tabular}{@{}l@{}}\begin{tabular}{@{}l@{}}S1: Did \{first\_name\} \{verb[0]\} the \{obj\}?\\S2: Was \{first\_name\} \{verb[1]\} by the \{obj\}?\end{tabular}\\\begin{tabular}{@{}l@{}}S1: Did Sam remember the factory?\\S2: Was Sam remembered by the factory?\end{tabular}\end{tabular} \\ \midrule
			Taxonomy &   Resolving external taxonomic hierarchy         &                                                                                                                                   \begin{tabular}{@{}l@{}}\begin{tabular}{@{}l@{}}S1: How can I become \{a:x[1]\} person?\\S2: How can I become \{a:x[0]\} person?\end{tabular}\\\begin{tabular}{@{}l@{}}S1: How can I become a frightened person?\\S2: How can I become a scared person?\end{tabular}\end{tabular} \\ \midrule
			Temporal &   Resolving temporal changes           &                                                                                                         \begin{tabular}{@{}l@{}}\begin{tabular}{@{}l@{}}S1: Is \{first\_name\} \{last\_name\} \{a:noun\}?\\S2: Did \{first\_name\} \{last\_name\} use to be \{a:noun\}?\end{tabular}\\\begin{tabular}{@{}l@{}}S1: Is Dorothy Clarke an agent?\\S2: Did Dorothy Clarke use to be an agent?\end{tabular}\end{tabular} \\ \midrule
			Vocabulary &      Resolving word choice variations       &                                                                                                                                                       \begin{tabular}{@{}l@{}}\begin{tabular}{@{}l@{}}S1: How can I become less \{x[1]\}?\\S2: How can I become more \{x[1]\}?\end{tabular}\\\begin{tabular}{@{}l@{}}S1: How can I become less active?\\S2: How can I become more active?\end{tabular}\end{tabular} \\ 
			
			\midrule
			\multicolumn{3}{c}{Natural Language Inference}\\
			\midrule
			Causal & \makecell[l]{Resolving actions between one\\ object and two entities}            &                            \begin{tabular}{@{}l@{}}\begin{tabular}{@{}l@{}}P: \{NAME\} moved the \{OBJECT\_TABLE\} to \{LOCATION\_HOUSE1\} from \{LOCATION\_HOUSE2\}.\\H: The \{OBJECT\_TABLE\} is now in \{LOCATION\_HOUSE1\}.\end{tabular}\\\begin{tabular}{@{}l@{}}P: George moved the notebook to study room from bedroom.\\H: The notebook is now in study room.\end{tabular}\end{tabular} \\ \midrule
			Conditional &  Resolving reasoning over conditions           & \begin{tabular}{@{}l@{}}\begin{tabular}{@{}l@{}}P: If \{NAME1\} comes to the \{LOCATION\}, \{NAME2\} won't come. \{NAME1\} did not come to the \{LOCATION\}.\\H: \{NAME2\} didn't come to the \{LOCATION\}.\end{tabular}\\\begin{tabular}{@{}l@{}}P: If Kim comes to the park, William won't come. Kim did not come to the park.\\H: William didn't come to the park.\end{tabular}\end{tabular} \\ \midrule
			Lexical &  \makecell[l]{Resolving word choice variations}           &                                                                                                                                        \begin{tabular}{@{}l@{}}\begin{tabular}{@{}l@{}}P: \{NAME\} was born in \{COUNTRY1\}.\\H: \{NAME\} is born in \{COUNTRY2\}.\end{tabular}\\\begin{tabular}{@{}l@{}}P: Emily was born in Germany.\\H: Emily is born in Malaysia.\end{tabular}\end{tabular} \\ \midrule
			Presupposition &  Resolving implications           &                                                                                                        \begin{tabular}{@{}l@{}}\begin{tabular}{@{}l@{}}P: \{NAME\}'s \{T12\_RELATION\} is \{ADJECTIVE\_OF\_PERSON\}.\\H: \{NAME\} has \{A\} \{T12\_RELATION\}.\end{tabular}\\\begin{tabular}{@{}l@{}}P: Florence's brother is intolerant.\\H: Florence has a brother.\end{tabular}\end{tabular} \\ \midrule
			Quantifier &   \makecell[l]{Resolving "all" (universal quantification) and\\ 
				"some", "none" (existential quantification)} &                                                                                 \begin{tabular}{@{}l@{}}\begin{tabular}{@{}l@{}}P: None of the \{OBJECTS\} are \{COLOUR1\} in colour.\\H: Some of the \{OBJECTS\} are \{COLOUR2\} in colour.\end{tabular}\\\begin{tabular}{@{}l@{}}P: None of the cars are maroon in colour.\\H: Some of the cars are pink in colour.\end{tabular}\end{tabular} \\ \midrule
			Syntactic &  Resolving ellipsis           &                                                                                                                         \begin{tabular}{@{}l@{}}\begin{tabular}{@{}l@{}}P: \{NAME\} tried but wasn't able to \{VERB\}.\\H: \{NAME\} didn't try to \{VERB\}.\end{tabular}\\\begin{tabular}{@{}l@{}}P: Alan tried but wasn't able to give.\\H: Alan didn't try to give.\end{tabular}\end{tabular} \\ \midrule
			World & Resolving world knowledge such as geography         &                                                                                                                                         \begin{tabular}{@{}l@{}}\begin{tabular}{@{}l@{}}P: \{NAME\} lives in \{T4\_CAPITAL1\}.\\H: \{NAME\} lives in \{T4\_COUNTRY2\}.\end{tabular}\\\begin{tabular}{@{}l@{}}P: Ken lives in Kathmandu.\\H: Ken lives in North Korea.\end{tabular}\end{tabular} \\  
			\bottomrule
	\end{tabular}}
\end{table*}

\begin{table*}
	\caption{The fine-tuned models we evaluated in this paper.}\label{tab:model}
	\centering
	\resizebox{\textwidth}{!}{
		\begin{tabular}{llll}
			\toprule
			Model name & Task & Size &                                                Checkpoint Identifier \\
			\midrule
			$\mathrm{DistillBERT}$ &  Sentiment Analysis & Small &           \ttt{textattack/distilbert-base-cased-SST-2} \\
			$\mathrm{ALBERT}$ &  Sentiment Analysis & Small &                       \ttt{textattack/albert-base-v2-SST-2} \\
			$\mathrm{BERT}_\mathrm{Base}$ &  Sentiment Analysis & Base&                   \ttt{textattack/bert-base-uncased-SST-2} \\
			$\mathrm{RoBERTa}_\mathrm{Base}$ &  Sentiment Analysis & Base&                        \ttt{textattack/roberta-base-SST-2} \\
			$\mathrm{DistillBERT}$ &  Paraphrase Detection & Small&                 \ttt{textattack/distilbert-base-cased-QQP} \\
			$\mathrm{ALBERTA}$&   Paraphrase Detection & Small &                        \ttt{textattack/albert-base-v2-QQP} \\
			$\mathrm{BERT}_\mathrm{Base}$ &   Paraphrase Detection & Base&                     \ttt{textattack/bert-base-uncased-QQP} \\
			$\mathrm{DistillBERT}$ &  Natural Language Inference & Small&                \ttt{textattack/distilbert-base-cased-snli} \\
			$\mathrm{ALBERT}$ &  Natural Language Inference  & Small&                       \ttt{textattack/albert-base-v2-snli} \\
			$\mathrm{RoBERTa}_\mathrm{Large}$ &  Natural Language Inference  & Large& \ttt{ynie/roberta-large-snli\_mnli\_fever\_anli\_R1\_R2\_R3-nli} \\
			\bottomrule
		\end{tabular}
	}
\end{table*}


\begin{table*}
	\caption{Proportion of valid sentences and performance of trained classifiers for automatic filtering.}
	\label{tab:acc}
	\resizebox{\textwidth}{!}{
		\begin{tabular}{lllccc}
			\toprule
			\makecell[l]{Linguistic\\ Capability} &                                                                            Test &  Accuracy &        F1 &  \makecell[c]{Proportion of\\ Valid Test Cases} \\ 
			\midrule
			\multicolumn{5}{c}{Sentiment Analysis}\\
			\midrule

			SRL &             A negative sentiment sentence with negative sentiment question and word yes as the answer. &         / &     / &     1.000 \\ \midrule
			SRL &             A positive sentiment sentence with positive sentiment question and word yes as the answer. &         / &     / &     0.994 \\ \midrule
			SRL &                          My opinion is more important than others' when expressing positive sentiment. &         / &     / &     0.948 \\ \midrule
			SRL &              A negative sentiment sentence with positive sentiment question and word no as the answer. &         / &     / &     0.928 \\ \midrule
			Temporal &               I used to have negative sentiment to something, but now I have positive sentiment to it. &         / &     / &     0.922 \\ \midrule
			Negation &                                              A negative sentiment sentence with negated positive word. &         / &     / &     0.910 \\ \midrule
			SRL &                          My opinion is more important than others' when expressing negative sentiment. &         / &     / &     0.900 \\ \midrule
			Temporal &               I used to have positive sentiment to something, but now I have negative sentiment to it. &     0.942 & 0.967 &     0.873 \\ \midrule
			Negation &                                                 I thought something was positive, but it was negative. &     0.934 & 0.961 &     0.860 \\ \midrule
			Vocabulary &                                                     A negative sentiment sentence with negative words. &     0.897 & 0.932 &     0.804 \\ \midrule
			Negation & A negative sentiment sentence with negated positive sentiment word and neutral contents in the middle. &     0.915 & 0.935 &     0.653 \\ 
			\midrule
			\multicolumn{5}{c}{Paraphrase Detection}\\
			\midrule
			
			Temporal &                             Two sentences are different when talking about a person's current job and his or her previous job. &         / &     / &     1.000 \\ \midrule
			Negation &                                     Two sentences are different when talking about someone should and should not do something. &         / &     / &     0.973 \\ \midrule
			Temporal &                               Two sentences are different when talking about a person's current job and his or her future job. &         / &     / &     0.964 \\ \midrule
			Vocabulary &                                                     Two sentences are different when adjectives are modified by more and less. &         / &     / &     0.912 \\ \midrule
			Taxonomy &                                             Two sentences are equivalent when the nouns are modified by synonymous adjectives. &     0.889 & 0.933 &     0.885 \\ \midrule
			Coref &                                                            Two sentences are different when swapping the subjects and objects. &     0.980 & 0.989 &     0.843 \\ \midrule
			Temporal &                                 Two sentences are different when talking about doing something before and after another thing. &     0.982 & 0.989 &     0.800 \\ \midrule
			Temporal &                               Two sentences are different when describing doing something before and after some specific time. &     0.982 & 0.988 &     0.782 \\ \midrule
			Negation &                                 Two sentences are different when talking about the properties of doing or not doing something. &     0.965 & 0.977 &     0.754 \\ \midrule
			Vocabulary &                             A sentence with the noun modified by an adjective is equivalent to the sentence without adjective. &     0.920 & 0.946 &     0.740 \\ \midrule
			SRL &                                                                          Two sentences are equivalent when using according to. &     0.958 & 0.971 &     0.708 \\ \midrule
			Negation & Two sentences are different when describing a person with adjective and a clause including the negation of the same adjective. &     0.871 & 0.913 &     0.690 \\ \midrule
			Taxonomy &     Two sentences are equivalent when one has an adjective modified by more and the other one has an antonym modified by less. &     0.920 & 0.909 &     0.511 \\ \midrule
			Coref &                                          Two sentences are different when referring someone's family using different pronouns. &     0.940 & 0.940 &     0.500 \\ \midrule
			SRL &                                                           Two sentences are different when swapping active and passive action. &     0.930 & 0.933 &     0.471 \\
			
			\midrule
			\multicolumn{5}{c}{Natural Language Inference}\\
			\midrule
			Lexical & \makecell[c]{/} &        / &  / &  0.975 \\ \midrule
			Syntactic & \makecell[c]{/} &        / &  / &  0.936 \\ \midrule
			Presupposition & \makecell[c]{/} &        / &  / &  0.903 \\ \midrule
			World & \makecell[c]{/} &        / &  / &  0.902 \\ \midrule
			Quantifier & \makecell[c]{/} &        / &  / &  0.895 \\ \midrule
			Causal & \makecell[c]{/} &        / &  / &  0.857 \\ \midrule
			Conditional & \makecell[c]{/} &        / &  / &  0.834\\
			
			\bottomrule
	\end{tabular}}
\end{table*}


\begin{table*}
	\centering
	\caption{Creating new templates based on test cases generated by GPT-3.}
	\label{tab:expansion}
	\begin{tabular}{c}
		\includegraphics[width=\textwidth]{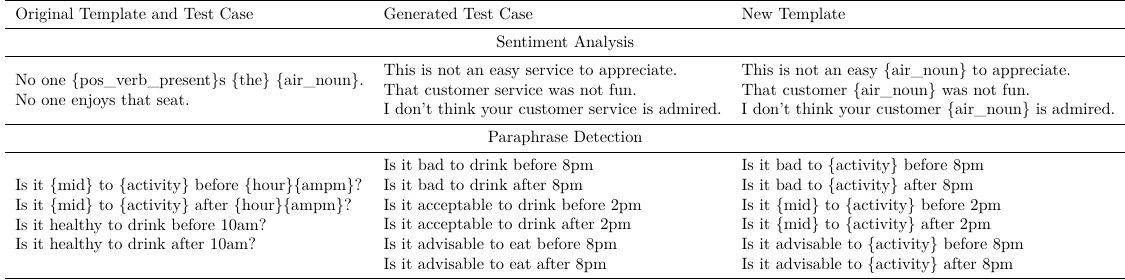}
	\end{tabular}
\end{table*}





\end{document}